\documentstyle[]{mn}
\newif\ifAMStwofonts
\def\BGE{\begin{equation}}
\def\EDE{\end{equation}}
\input epsf
\ifoldfss
  \ifCUPmtlplainloaded \else
    \NewTextAlphabet{textbfit} {cmbxti10} {}
    \NewTextAlphabet{textbfss} {cmssbx10} {}
    \NewMathAlphabet{mathbfit} {cmbxti10} {} % for math mode
    \NewMathAlphabet{mathbfss} {cmssbx10} {} %  "   "    "
  \fi
  \ifAMStwofonts
    \ifCUPmtlplainloaded \else
      \NewSymbolFont{upmath} {eurm10}
      \NewSymbolFont{AMSa} {msam10}
      \NewMathSymbol{\upi}     {0}{upmath}{19}
      \NewMathSymbol{\umu}     {0}{upmath}{16}
      \NewMathSymbol{\upartial}{0}{upmath}{40}
      \NewMathSymbol{\leqslant}{3}{AMSa}{36}
      \NewMathSymbol{\geqslant}{3}{AMSa}{3E}

      \let\leq=\leqslant 
      \let\geq=\geqslant 
    \fi
  \fi
\fi % End of OFSS

\ifnfssone
  \newmathalphabet{\mathit}
  \addtoversion{normal}{\mathit}{cmr}{m}{it}
  \addtoversion{bold}{\mathit}{cmr}{bx}{it}
  \newmathalphabet{\mathbfit} % math mode version of \textbfit{..}
  \addtoversion{normal}{\mathbfit}{cmr}{bx}{it}
  \addtoversion{bold}{\mathbfit}{cmr}{bx}{it}
  \newmathalphabet{\mathbfss} % math mode version of \textbfss{..}
  \addtoversion{normal}{\mathbfss}{cmss}{bx}{n}
  \addtoversion{bold}{\mathbfss}{cmss}{bx}{n}
  \ifAMStwofonts
    \ifCUPmtlplainloaded \else
      \UseAMStwoboldmath
      \makeatletter
      \new@mathgroup\upmath@group
      \define@mathgroup\mv@normal\upmath@group{eur}{m}{n}
      \define@mathgroup\mv@bold\upmath@group{eur}{b}{n}
      \edef\UPM{\hexnumber\upmath@group}
      \new@mathgroup\amsa@group
      \define@mathgroup\mv@normal\amsa@group{msa}{m}{n}
      \define@mathgroup\mv@bold\amsa@group{msa}{m}{n}
      \edef\AMSa{\hexnumber\amsa@group}
      \makeatother
      \mathchardef\upi="0\UPM19
      \mathchardef\umu="0\UPM16
      \mathchardef\upartial="0\UPM40
      \mathchardef\leqslant="3\AMSa36
      \mathchardef\geqslant="3\AMSa3E

      \let\leq=\leqslant 
      \let\geq=\geqslant 
    \fi
  \fi
\fi % End of NFSS release 1

\ifnfsstwo
  \DeclareMathAlphabet{\mathbfit}{OT1}{cmr}{bx}{it}
  \SetMathAlphabet\mathbfit{bold}{OT1}{cmr}{bx}{it}
  \DeclareMathAlphabet{\mathbfss}{OT1}{cmss}{bx}{n}
  \SetMathAlphabet\mathbfss{bold}{OT1}{cmss}{bx}{n}
  \ifAMStwofonts
    \ifCUPmtlplainloaded \else
      \DeclareSymbolFont{UPM}{U}{eur}{m}{n}
      \SetSymbolFont{UPM}{bold}{U}{eur}{b}{n}
      \DeclareSymbolFont{AMSa}{U}{msa}{m}{n}
      \DeclareMathSymbol{\upi}{0}{UPM}{"19}
      \DeclareMathSymbol{\umu}{0}{UPM}{"16}
      \DeclareMathSymbol{\upartial}{0}{UPM}{"40}
      \DeclareMathSymbol{\leqslant}{3}{AMSa}{"36}
      \DeclareMathSymbol{\geqslant}{3}{AMSa}{"3E}

      \let\leq=\leqslant 
      \let\geq=\geqslant 
    \fi
  \fi
\fi % End of NFSS release 2

\ifCUPmtlplainloaded \else
  \ifAMStwofonts \else % If no AMS fonts
    \def\upi{\pi}
    \def\umu{\mu}
    \def\upartial{\partial}
  \fi
\fi

\title{Evolution of the Globular Cluster System in a Triaxial Galaxy}
\author[R.~Capuzzo--Dolcetta and A.~Tesseri]
       {R.~Capuzzo--Dolcetta,$^1$  A.~Tesseri,$^1$\\
        $^1$ Istituto Astronomico, Universit\`a La Sapienza,\\
 Via G. M. Lancisi 29, I-00161, Roma, Italy\\
dolcetta@astrmb.rm.astro.it,
tesseri@astrmb.rm.astro.it}
       
\date{Accepted  yr m d.
      Received yr m d;
      in original form yr m d}

\pagerange{\pageref{firstpage}--\pageref{lastpage}}
\pubyear{yr}

\begin{document}

\maketitle

\label{firstpage}

\begin{abstract}
Dynamical friction and tidal disruption are effective mechanisms of
evolution of globular cluster systems, especially in non--axysimmetric
galaxies with a central compact nucleus. With a semi--analytical approach based on the knowledge
of the dependence of the dynamical friction and tidal disruption 
effects on the relevant parameters, we are able to follow the
time evolution of the globular cluster system in a model of a
triaxial galaxy and give its observable properties to compare
with observational data.

An important result is that the flatter distribution of the globular cluster system relatively to that of the stellar bulge, as observed in many galaxies, can be explained by the evolution of the globular cluster
system, starting from the same density profile.
\end{abstract}


\section{Introduction}
Two observational facts are well established, by now:

$i)$ the first is that many 
galaxies have globular clusters systems (GCSs) with density profiles less 
concentrated than their parent galaxy halo light, M87 and M49 being the prototypes (Lauer \& Kormendy 1986, Harris 1986,1991).
This has been recently confirmed by observations of a sample of 14 elliptical galaxies, made with the WFPC2 of the Hubble Space Telescope (Forbes et al. 1996).
These observations, thanks to the high resolution of the HST, were able to probe the inner kiloparsecs of those galaxies and show that most, if not all, of them have GCSs with surface density profiles
 that rise towards the centre less steeply than the underlying galaxy light;

$ii)$ the second is that many galaxies host compact massive nuclei in their 
centres (Dressler \& Richstone 1988, Kormendy 1988, Kormendy \& Richstone 
1995, Eckart \& Genzel 1996) with estimated masses in the range from 2$\cdot 10^6
 M_\odot$ for our Galaxy and M32 up to 3$\cdot 10^9 M_\odot$ 
for M87.

An explanation for the difference between halo star and globular 
cluster distributions has been proposed by Harris \& Racine (1979), Harris 
(1986), and Racine (1991) as a difference in the formation epoch of the 
two components. In this scenario, the GCS formed in an earlier phase of the 
protogalactic collapse, while the stars that constitute the halo condensed 
later, this resulting in a less peaked distribution for the clusters. 
This picture requires an exact timing in the sequence of the evolution, in 
order to permit the clusters to be less metal 
rich than the halo stars, while producing the required differences in central 
concentration. In disk galaxies,however, 'the epoch of cluster formation would be early
 enough to force chenical enrichment but not early enough to take on a distinct spatial structure' (Harris 1986, Sect. VII, p. 840). Moreover, this scenario does not explain why the tails of the two density distributions are almost the same.
This last observational evidence suggests an alternative explanation: the cluster system and 
the halo formed at the same time with a similar spatial distribution, and the present 
differences are a consequence of the dynamical evolution of the GCS. Dynamical evolution correlates also with the possible presence of masssive central nuclei.
Actually, the larger core radius of the GCS would imply that the globular 
cluster population has been significantly depauperated in the inner regions of 
the system. This is the case when a massive object (like a black hole) resides 
in the centre of a galaxy and disrupts, by means of tidal forces, 
globular clusters which pass sufficiently close to it. There is no direct 
evidence that the aforementioned massive objects are black holes (they could 
be, as proposed by Kormendy \& Richstone (1995), massive clusters of low-mass 
stars, stellar remnants etc.) except in the case of NGC 4258, 
where the discovery of a perfect keplerian rotation curve in the inner regions 
(Miyoshi et al. 1995) rules out, on dynamical evidences (Mayoz 1995), 
alternatives to black holes. In the outer regions, on the contrary, 
the influence of the central massive object will be negligible. So we expect 
that there the  cluster distribution has remained more or less unchanged. 

Another major dynamical effect, the dynamical friction of field stars on globular clusters, enhances the efficiency of the depleting mechanism. It acts 
reducing the cluster orbital energy and bringing the clusters towards the centre, thus increasing the number of globular clusters in the inner regions. These clusters could feed the massive object. This scenario has been proposed first by Tremaine, Ostriker and Spitzer (1975), in a study on M31, where they showed 
how a massive object of mass in the range 10$^7$--10$^8$ $M_\odot$ could 
directly form from globular clusters braked to the centre of the galaxy and 
there merged.

Both of these effects (dynamical friction and tidal disruption) are 
significantly emphasized if the galaxy is triaxial in shape, 
a possibility supported by several observations (see, for example, Bertola, 
Vietri \& Zeilinger (1991) which show evidence of triaxial distributions in 
32 galaxies). The orbits which constitute the bulk of such a potential are the 'box' orbits, which are dense around the centre (see, e. g., Binney \& Tremaine 1987, hereafter BT) where the massive object lies (so that even globular 
clusters of large apocentric distance will possibly be disrupted) and the field star density is higher (so that dynamical friction efficiency is 
increased). In fact, Pesce, Capuzzo Dolcetta and Vietri (1992) have 
demonstrated that dynamical friction decay times on box orbits are significantly 
reduced.

The shape of the velocity ellipsoid of halo stars and globular clusters in our galaxy supports this picture. In fact, while the velocity dispersion 
of the halo stars is larger in the radial direction, as expected from 
numerical simulation of the radial collapse of the protogalaxy, the globular 
clusters' velocity ellipsoid is almost spherical. Under the hypothesis of a 
coeval formation, this can be explained by a selective process which destroyed 
the clusters on low-angular momentum orbits (i.e. box orbits in a triaxial galaxy). The role of triaxiality in the tidal disruption mechanism has been quantitatively discussed by Ostriker, Binney and Saha (1989) (hereafter OBS). In their paper they proposed, for the first time, that the formation of a massive nucleus from decayed globular clusters could be a self-limiting process, due to the inverse proportionality of the tidal disruption 
timescale, $\tau_{tid}$, to the nucleus mass. Capuzzo Dolcetta (1993) has investigated thoroughly the evolution of a GCS in 
the Schwarzschild's (1979) triaxial non-rotating model, under the combined 
effects of dynamical friction and tidal disruption, studying the growth of the 
nucleus and the evolution of globular clusters' mass function. 
Indeed, he found that the cooperation of these two effects may lead to the 
formation of a compact nucleus, in form of globular clusters decayed to the 
centre of the galaxy. The growth of the nucleus eventually 
halts when its mass is large enough to shatter all incoming clusters. Of 
course the value of the mass reached by the growing nucleus depends, in this 
scheme, on the initial GCS spatial, mass and velocity distributions.

The increased efficiency of dynamical friction and tidal disruption, in a 
triaxial galaxy, is so increased that their effects are not limited to the 
very inner regions of the parent galaxy. Moreover, there is no need to assume, 
for the GCS, a box-biased phase-space density such that the 
globular clusters are all on box orbits. In fact, Capuzzo
Dolcetta (1993) showed that even in the case of an isotropic distribution function (hereafter DF), the evolution of the GCS is very similar to that of a box-biased 
DF. The reason for this is that the two DFs do not differ much in the region of 
the phase-space occupied by the majority of the clusters.

In Section 2 we describe our model of a globular cluster system evolving
due to dynamical friction and tidal disruption effects and we give a formula
which permits to calculate the density profile of the GCS and its
observable properties; in Section 3 we present and discuss  the results. 

\section{The model}
In this paper we study a population of clusters exclusively on box orbits. 
This may be the case if they formed in the early galactic stages, during the radial collapse of the protogalactic nebula (van Albada 1982, Binney 1988).
We develop a semi-analytical model which allows to follow the evolution of the 
spatial density of the GCS in a triaxial galaxy under the influence of the two 
main evolutionary effects: dynamical friction and tidal disruption. Once the spatial density profile of the GCS is obtained, we can deduce its surface 
density profile, core radius and other useful quantities which may be compared 
to observations.
The galactic potential adopted here, like in Pesce et al. (1992) 
and Capuzzo-Dolcetta (1993), is that of the Schwarzschild's (1979) model. 
We define (following Capuzzo Dolcetta 1993) $\tau_{df}(E,m)$ as the time required to a cluster on a box orbit to lose all its energy $E$ (and stop at the centre of the galaxy) by means of the frictional drag exerted by the stellar population. A good fit to $\tau_{df}(E,m)$ is:
\BGE
   \tau_{df}(E,m)={1 \over m} {7.5 \cdot 10^8 \over (1-E)^2}~yr,
\EDE	
where $m$ is the mass of the cluster in units of 10$^6 M_\odot$ and E (0$\leq E <$ 1) is the orbital energy per unit mass in units of $\Phi_0$,  the central value of the Schwarzschild potential. Equation (1) is not well behaved at the origin since, as the energy approaches zero, $\tau_{df}(E,m)$ reaches a finit limit, which is inconsistent with its definition. In practical calculations we used a slightly different form of Eq. (1), which exhibits  the behavior required.

To include the effect of tidal disruption we will need an estimate of the 
timescale, $\tau_{tid}$, associated to this effect which, according to OBS and Capuzzo Dolcetta (1993), may be expressed in the form
\BGE
\tau_{tid}= {1 \over \mu\pi\sqrt{5}} \sqrt{Gm \over R_h}{v_n\over v_c^2}
 {A_w\over r_c R_h} T_r
\EDE			
where: $\mu = {GM_n\over r_c v_c^2}$; M$_n$ is the mass of the nucleus; r$_c$ is the core radius
of the galaxy;  v$_c$ is the circular speed at large r; v$_n$ is the speed of the cluster of mass $m$ at the point where the gravitational attraction of the ellipsoid equals that of the nucleus; $A_w$ is the area of the waist of the box orbit; $R_h$ is the half-mass radius of the cluster; $T_r$ is the half-period of oscillation parallel to the potential long axis.
Note that Eq. (2) depends on the structural parameters of the clusters only 
through the quantity $\sqrt {m \over R_h^3}$ which is proportional to the square root of the mean density within the half-mass radius, $\sqrt {\bar \rho_h}$. The dependence of $\tau_{tid}$ on the orbital energy $E$ is through $A_w$, $v_n$ and $T_r$.

Since we are interested in the density profiles of the GCS, it is useful to treat it as a `fluid' system. This allows us to adopt the continuity equation as the one which rules the evolution of the system:
\BGE
{\partial \rho \over \partial t} + {\partial \rho v_r \over \partial r}
+ 2 {\rho v_r \over r} = -S
\EDE
where $\rho(r,t)$ is the number density of the GCS, $v_r(r,t)$ is its radial 
velocity field and S $\geq 0$ is a `sink' term. We have written the continuity 
equation in spherical coordinates, dropping the terms containing the 
derivatives with respect to angles, since the observed GCSs usually show 
spherical symmetry. Moreover, we have adopted $v_\theta$=0 and $v_\phi$=0 since the GCSs seem not to rotate.
If no evolution of the GCS occurs (as is the case when the system is collisionless and no massive central `absorber' is present), we expect 
$v_r$(r,t)=0 and S=0 for any $r$ and $t$. In our picture, instead, the frictional drag of the stellar halo on 
the GCS, reduces the energy of the clusters, acting as a radial velocity field pointing inward. At the same time, the massive galactic nucleus erodes the GCS, standing for a non-zero sink term 
$S=\rho/\tau_{tid}$, where, hereafter, $\tau_{tid}$, is the average of formula (2) over the DF of the system, as described in Appendix A. Once we give the initial GCS density 
distribution $\rho(r,0)$, and we assume $v_r(r,t)$=$v_r(r)$ 
(that is, the radial velocity field generated by the unevolving stellar halo does not change during the evolution of the system), the following solution of Eq. (3) is obtained (see Appendix B):
\BGE
\matrix{
& \rho(r,t)&=&{v_r\left({ H(r,t) }\right) H^2(r,t) \over v_r(r)r^2} \rho 
\left( H(r,t),0 \right) \times \cr
& \cr
& &\times& \exp\left({-\int_0^t{dx \over \tau_{tid}\left(H(r,x),M_n(t-x),\bar \rho_h
\right)} }\right)
}
\EDE
where 
\BGE
H(r,t)=H\left( t+\tau(r) \right)
\EDE
\BGE
\tau(r)=- \int {dr \over {v_r(r)}}
\EDE 
and H(x) is the inverse function of $\tau$(r). 
The function M$_n$(t), which gives the nucleus mass at any time t, is not known. In our calculations, we will adopt a fixed nucleus mass since a reliable evaluation 
of the rate of accretion of the nucleus requires further investigation 
(see Capuzzo-Dolcetta 1996). Anyway, when a realistic model will be available to give M$_n$(t), it will be 
straightforwardly included in our model.

By the definition of $\tau$(r), the difference $\tau$(R) - $\tau$(0) is the 
time required to an element of mass of the `fluid' to reach the centre and stop 
there, starting from a point at distance R from the origin. It is so 
straightforward to relate this $\tau$(r) to the function $\tau_{df}$(E,m). This is easily done by averaging $\tau_{df}(E,m)$ over the DF of the system, thus obtaining a function $\tau_{df}$(r,m) (see Appendix B).
Now, the velocity field at any point r can be estimated by
\BGE
v_r(r,m)= -{r \over {\tau_{df}}}.
\EDE
Thus, substituting Eq. (7) in Eq. (6), we find the following expression for the 
function $\tau$(r) which appears in the solution (4):
\BGE
\tau(r,m)= \int{\tau_{df}\over r} dr.
\EDE
In this way we have related the dynamical friction effect to the presence of a velocity field pointing radially inward. As expected, $\rho(r,t)$ as given by Eq. (4), depends on $m$ via Eq. (8). In the following, we assume the initial distribution of the GCS in the form
\BGE
\rho(r,0;m)=\psi_0(m) \cdot \rho(r,0)
\EDE
where $m$ is the individual cluster mass and $\psi_0$(m) is the GCS's initial mass function (hereafter IMF). The relation (9) says that the IMF of the GCS  is indipendent of 
the position in the system. The solution of eq. (3), in this case, is given by a 
superposition of distributions:
\BGE
\rho(r,t)=\int_{m_1}^{m_2} \rho(r,t;m) \psi(m) dm
\EDE
where $\rho$(r,t;m) is the solution (4) for a single mass GCS, taking into 
account the dependence of $\tau_{df}$ on the clusters' mass, m. To conclude, the dependence on the 
overall model is given by the functions $\tau_{df}$ and $\tau_{tid}$, and by the 
GCS's DF which fixes its initial distribution. It is clear, hence, that equations (4) and (10) may represent the evolution of a GCS in different situations, provided we know $\tau_{df}$, $\tau_{tid}$ and $f(E)$.

\section[]{The results}
In the previous Section we have solved the  evolution equation for the density distribution of a GCS made up of clusters of different masses when dynamical friction and tidal disruption are taken into account. Before showing the results, we specify the parameters which our model depends on.

For $\bar \rho_h$, we seek a relation with the cluster mass. From data 
of clusters in our Galaxy (Webbink 1985) we found a loose relation among the 
mass and the half-mass density of the clusters, obtained by adopting a fixed 
ratio $(M/L_V)_\odot$=1.5 and fitting the clusters density 
profiles with Plummer models:
\BGE
\bar \rho_h=1.38 \cdot 10^5~ m^{1.56}~M_\odot pc^{-3}
\EDE
This choice for the cluster half-mass density is quite different from that 
used by OBS in their calculations. Their GCS is composed of identical 
clusters with half-mass density $\bar \rho_h=10^3$ M$_\odot$pc$^{-3}$, implying that we are dealing with clusters significantly denser than theirs. Since the tidal disruption timescale depends on $\sqrt{\bar \rho_h}$, we expect a lesser depletion of the GCS.

As initial single-mass number density for the GCS we assume:
\BGE
\rho(r,0) = {\rho_0\over \left[{1+\left(r\over r_{c0} \right)^2}\right]^{3 \over 2}}
\EDE
which corresponds to the monopole component of the Schwarzschild potential, 
$\rho_0$ being the central density and $r_{c0}$ the initial core radius (assumed equal to the bulge star core. 
Note that the projection, $\Sigma$(r), of (12) is the so-called modified 
Hubble law, a function well fitting the elliptical galaxies' surface brightness:
\BGE
\Sigma(r,0)={\Sigma_0\over 1+\left(r \over r_{c0} \right)^2}
\EDE
where $\Sigma_0=2 \rho_0$r$_{c0}$. Since the density distribution (12) has an infinite
 mass we cut it at a radius r$_{max}\gg $r$_{c0}$. 
Varying the value of r$_{max}$ results in a different behaviour of the averaged 
timescales $\tau_{df}$ and $\tau_{tid}$. Anyway, we find that, letting r$_{max}$ to vary in a reasonable range, the differences are more important in the outer regions ( r $\geq$ 20~r$_{c0}$), where, anyway, the timescales are always sensibly longer than a Hubble time. In the regions of 
interest (r$\leq$ 10~r$_{c0}$), the changes in r$_{max}$ reflect in negligible changes in our timescales.

As DF for the GCS, we use a King model with $\sigma^2$=8$\Phi_0$ (see eq. 4-130 and 4-131 in BT), whose corresponding density profile fits well eq. (12), as required.

For the GCS IMF we shall assume truncated power-laws:
\BGE
\psi_0(m)=\left \lbrace \matrix{&0           &0<m<m_1 & \cr
                              &k~m^{-s} & m_1 \leq m \leq m_2 & \cr
                              &0              &m > m_2  &} \right.
\EDE		
\begin{figure}
\vspace{1pt}
\epsfxsize=250pt
\epsfbox{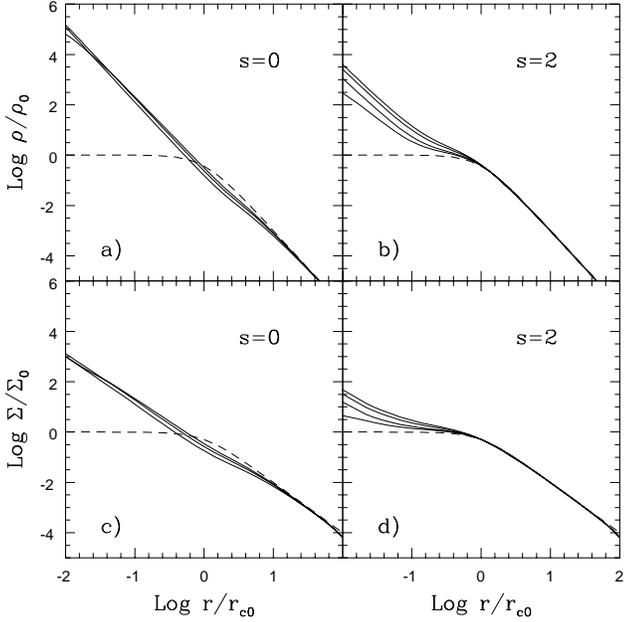} 
 \caption{Evolution of the GCS space and surface density (upper and lower 
 panels,respectively) due to dynamical friction only. Left panels: flat (s=0) IMF; right panels: steep (s=2) IMF. The thick curve is the initial 
distribution, the other curves correspond to 1, 5, 10 and 15 Gyr 
(bottom to top)}
\end{figure}
with $m_1=10^4 M_\odot$ and $m_2=3\cdot 10^6 M_\odot$, so that the product $\psi_0(m)\rho(r,0)$ gives the intial number per unit volume of clusters with mass $m$. We consider the two cases 
of a flat (s=0) and a steeply decreasing (s=2) IMF. The normalization factor $k \rho_0$ is chosen in such a way to give a total number of clusters
$N_{tot}$=1000.

\begin{figure}
\vspace{1pt}
\epsfxsize=250 pt
\epsfbox{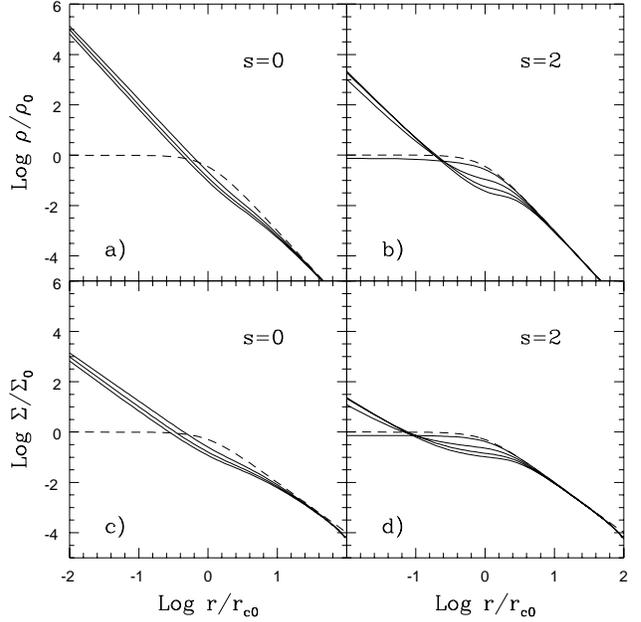}
\caption{Same as Fig.1, taking into account tidal disruption of clusters by
a nucleus of fixed mass M$_n$=10$^7$ M$_\odot$. The thick curve is the initial 
distribution, the curve corresponding to 1 Gyr is the closest in shape to the initial
distribution and the other curves correspond to 5, 10 and 15 Gyr (top to bottom)}
\end{figure} 

\subsection{Dynamical friction}
When a pre-existing nucleus is absent - or its mass is so small that the tidal disruption effect is negligible - we may apply formula (B.11) for the evolution of the number density distribution of a GCS undergoing 
dynamical friction only. The results are shown in Fig. 1. 
We observe an inner region (within r$\approx$ r$_{c0}$)where the density is strongly enhanced, surrounded by a depleted strip (r$_{c0} \leq$ r $\leq$ 10~r$_{c0}$). This effect is best seen in the s=0 case (Fig. 1a,c), 
since, with such an IMF, the GCS is composed by a large fraction of massive 
clusters, which decay faster. On the contrary, with a steep (s=2) IMF (Fig. 1b,d), the GCS is composed mainly of light clusters, which respond slowly to the frictional 
drag, thus the depletion in the outer region is almost imperceptible.
The surface distributions, in both cases, become more concentrated as time goes on, so that the core radii become smaller. This is a natural consequence of the fact that globular clusters lose their energy by dynamical friction and move on less extended orbits.

To study the time evolution of the core radius of the GCS, we need to remember that an actual observation cannot sample the GCS distribution all the way to the centre of the galaxy. As a consequence, the value of the observed core radius depends on a somewhat arbitrary 
extension of the GCS' surface density towards the centre of the galaxy. 
Were the inner limit of the observations equal to, say, $r_{c0}$, in the s=2 case we would not observe any differences between the halo and GCS distributions (Fig. 1d); in the s=0 case we would observe a flatter slope (Fig. 1c), that would lead us to conclude erroneously that the GCS is less concentrated than the halo.
Could we sample clusters to a smaller inner radius, say 0.1 r$_{c0}$, we would observe that the GCS istributions, in both cases, are more concentrated than that ofthe halo.
It is evident, then, the importance of going with the observations as close to the 
centre of the galaxy as possible. In the next Subsection we give a more detailed evaluation of the core radius of the evolved GCS. 

It is interesting to calculate the number of clusters contained in a given radius $R$ at time $t$, $N_{cl}(R,t)$. In the case of dynamical friction alone, the integral over r is easily done, and we obtain:
\BGE
N_{cl}(R,t)= \int_{m_1}^{m_2} {N_{cl}^0 \left( H \left(t + \tau_{df}(R,m)\right) \right) \psi_0(m) dm}
\EDE
where $N_{cl}^0$(r) is the number of clusters in the sphere of radius $r$ at $t=0$. A similar equation holds for the mass contained in the sphere of radius $R$ at 
time $t$, $M_{cl}(R,t)$. In this way, we can evaluate the number and mass of clusters within 
the core at present time ($t$=15 Gyr), $N_{cl}$ and $M_{cl}$, and compare them with 
their initial values, $N_{cl0}$ and $M_{cl0}$ (see Table 1).
\begin{table}
 \caption{Number and total mass of clusters inside one core radius at time 15 Gyr, 
 compared to their initial values (with the subscript 0).}
 \label{symbols}
 \begin{tabular}{@{}lcccc}
  IMF & N$_{cl0}$ & N$_{cl}$ & M$_{cl0}$ (M$_\odot$) & M$_{cl}$ (M$_\odot$) \\
\hline
  s=0 & 37 & 410 & 5.6$\cdot 10^7$ &  7.2$\cdot 10^8$ \\
  s=2 & 37 & 66 & 2.1$\cdot 10^6$ &  1.2$\cdot 10^7$ \\
\hline 
\end{tabular}
\end{table} 
This table gives a first insight on the possible formation of a massive nucleus by globular clusters accretion, since the M$_{cl}$ values constitute the upper limits to the mass contribution by the cluster system. 
It is easily seen that $N_{cl}(R,t)$, as well as $M_{cl}(R,t)$, scale linearly with $N_{tot}$. So, bigger values for the mass inside the core radius require larger $N_{tot}$, which is not a free parameter, being constrained by observations, through the presently observed number of clusters.

\subsection{Dynamical friction and tidal disruption}
Now, we show the results for the volume density and surface distribution of a GCS subjected to both dynamical friction and tidal disruption. As it is easilyseen, in Figs. 2, 3 and 4 the density profiles strongly depend on $M_n$

\begin{figure}
 \vspace{1pt}
\epsfxsize=240 pt
\epsfbox{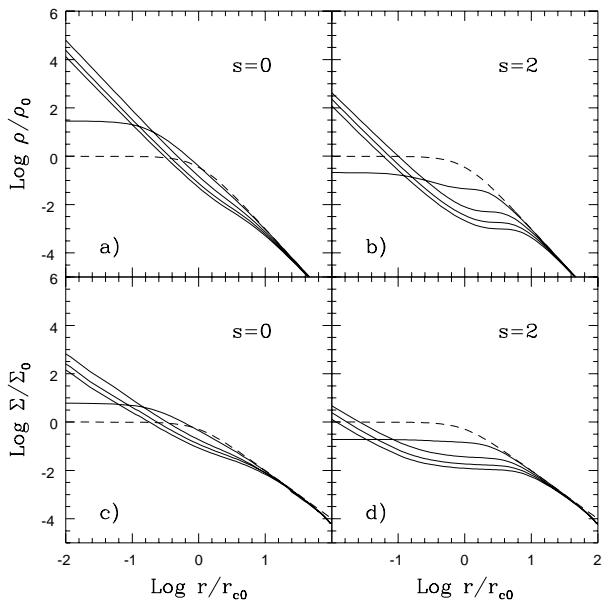} 
 \caption{Same as Fig.2, but M$_n$=10$^8$ M$_\odot$.}
\end{figure} 
\begin{figure}
 \vspace{1pt}
\epsfxsize=240 pt
\epsfbox{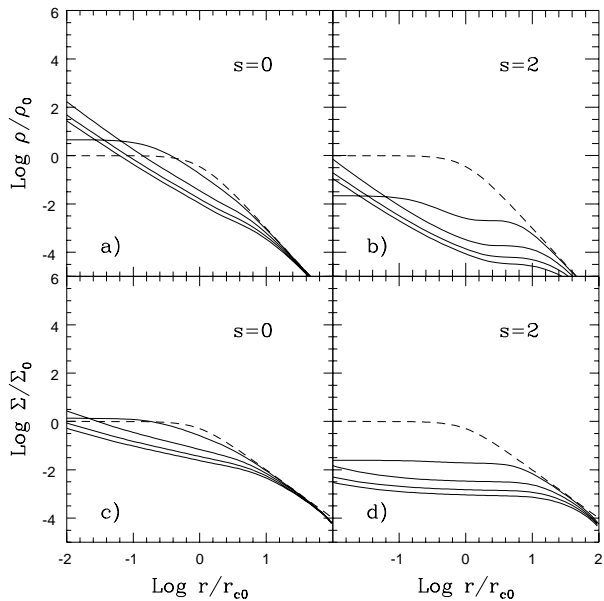} 
 \caption{Same as Fig.2, but $M_n$=10$^9$ $M_\odot$.}
\end{figure} 
The role of the nucleus is overwhelming when the GCS is composed mainly of low-density clusters (s=2 case, see Eq. (11)). In this case, the profiles differ significantly from the case of the absence of a nucleus, even if the nucleus mass is moderate ($M_n$=10$^7$ $M_\odot$, compare Fig. 1b,d to 2b,d). 
On the contrary, the more massive clusters of the flat IMF, being denser, resist effectively to the tidal disruption when $M_n \leq 10^7 M_\odot$ (compare Fig. 1a,c to 2a,c). Of course the depauperating effect of tidal disruption is enhanced by heavier nucleus masses (see Fig. 3 and 4).

Another interesting feature is the prominent central density cusp, more evident for light nuclei and flat IMF. This is due essentially to heavy clusters which rapidly move towards the centre of the galaxy due to large dynamical friction suffered by their high mass, and 
contemporarily survive tidal disruption because of their high density. 
At increasing nucleus masses, this effect is less evident and becomes almost imperceptible for heavy nuclei, because of the increasing efficiency of tidal disruption. Note how the differences among the density profiles of the innermost and external regions are significantly reduced by projection (compare upper and lower panels in Fig. 2,3 and 4).
\subsubsection{The GCS core radius evolution}
In the s=2 case for the IMF, the surface distributions depart from a modified Hubble law just in the inner regions, thus leading to a reliable r$_c$ determination. With a flat IMF, instead,  no region of constant density (a `core') is kept up to present time. Anyway, observationally, the steepening of the profile (which carries the signature of dynamical friction) could be appreciated just inside a region whose radius is of the order of $r_{c0}$. In this case, the evaluation of $r_c$ depends on the extension of the observed profile 
towards the centre, which usually results in its overestimate. 
\begin{figure}
 \vspace{1pt}
\epsfxsize=240 pt
\epsfbox{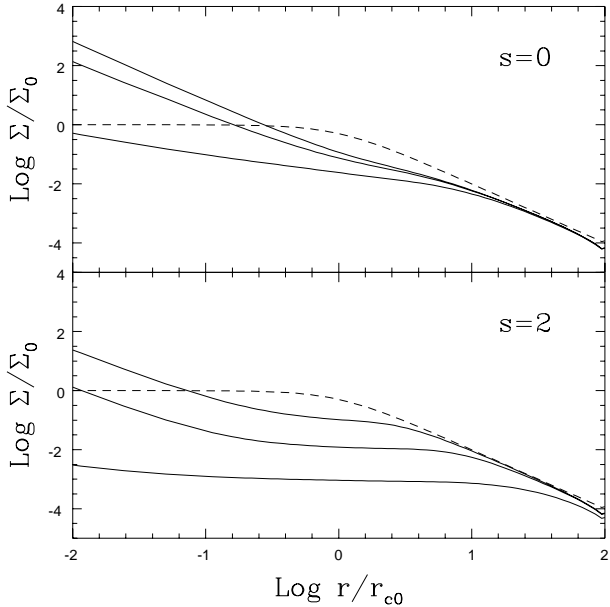} 
 \caption{Plot of the GCS surface density profile after evolution has occurred for a flat (s=0) IMF (upper panel) and for a steep (s=2) IMF (lower panel). The thick curve is the initial profile, the other curves correspond to nucleus masses of 10$^7$, 10$^8$ and 10$^9$ $M_\odot$ (top to bottom).}
\end{figure} 
To give a quantitative evaluation of this overestimate, we have calculated the 'observable' core radii of the GCS after a dynamical evolution of 15 Gyr, by fitting the surface density profiles of Fig. 5 with modified Hubble laws. By 'observable' we mean that the fits were done excluding too inner regions, within some radius $r_{min}$. The results are given in Tables 2 and 3, for different nucleus masses. In the s=2 case the value of $r_c$ does not change much on varying $r_{min}$, confirming that the surface distribution of the GCS is well approximated by a modified Hubble law. On the contrary, in the s=0 case, the observed core radius 
depends strongly on the value of r$_{min}$, indicating that the surface distribution does not display a well defined central core.
Then, a core is reliably defined when clusters have been destroyed by tidal disruption.
\begin{table}
\caption{`Observed' core radius for the case of a flat IMF (s=0).}
\label{symbols}
 \begin{tabular}{@{}lccc}
    r$_{min}$ & M$_n$=10$^7$ & M$_n$=10$^8$ & M$_n$=10$^9$  \\
\hline  
  r$_{c0}/2   $   & 2.7  & 8.7 & 24.0  \\
  r$_{c0}$        & 3.0  & 9.3 & 25.0  \\
  5 r$_{c0}$      & 3.0  & 9.3 & 25.0 \\
\hline
\end{tabular}

\medskip
Values obtained by fitting the surface density profiles of the evolved GCS
with a modified Hubble law, excluding the region inside r$_{min}$, for different
nucleus masses M$_n$.
\end{table}
\begin{table}
\caption{Same as Table 2, but for the case of a steep IMF (s=2).}
\label{symbols}
 \begin{tabular}{@{}lccc}
  r$_{min}$ & M$_n$=10$^7$ & M$_n$=10$^8$ & M$_n$=10$^9$  \\ 
\hline  
r$_{c0}$/2   & 0.8  & 0.8 & 1.6  \\
  r$_{c0}$        & 1.4  & 2.0 & 4.2  \\
  5 r$_{c0}$      & 4.6  & 5.2 & 7.5  \\
\hline 
\end{tabular}
\end{table}

In Fig. 6, we plot, for the steep IMF, the time evolution of the `observed' core radius for different nucleus masses, obtained as described above.
\begin{figure}
 \vspace{1pt}
\epsfxsize=240 pt
\epsfbox{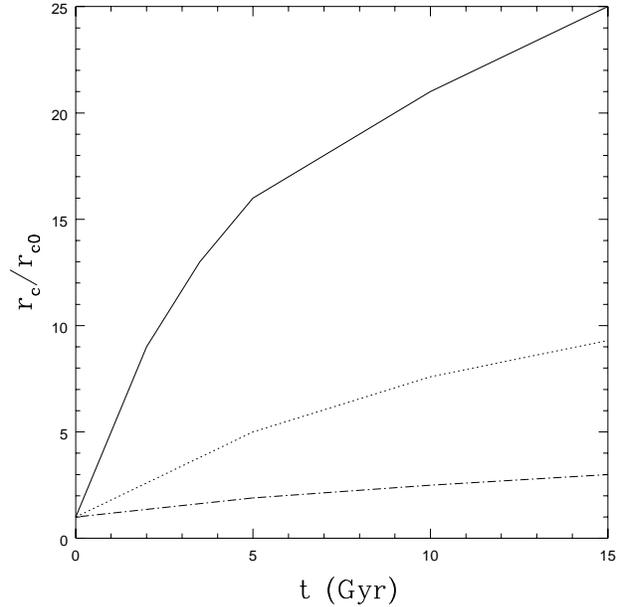} 
 \caption{Time evolution of the `observed' core radius for a GCS with a steep
 (s=2) IMF and $M_n$=10$^7$ $M_\odot$ (dot--dashed curve), $M_n$=10$^8$ $M_\odot$ (dotted curve),	$M_n$=10$^9$ $M_\odot$ (solid curve)}
\end{figure} 
\subsubsection{A parameter reliably comparable with observations}
A better parameter to synthesize the properties of the evolved GCSs is given by $x=N_l/N_0$, defined as the ratio between the number of `lost' clusters outside the minimum radius reachable by the observations (r$_{min}$) to the initial number of clusters in that region. The number $N_l$ is given by the difference between the initial and the present number of clusters, outside $r_{min}$. 
The `lost' clusters include either clusters which have been actually destroyed by tidal interaction with the massive nucleus either clusters which have lost energy by dynamical friction and have moved on orbits all within $r_{min}$.

The quantity $x$ is a good parameter to quantify the roles of the evolutionary mechanism under discussion, because it could be easily inferred from observations, as the difference observed between the (normalized) radial profiles of halo stars and GCS. This requires the reasonable hypothesis that the initial distributions of clusters and bulge stars were the same and that the present bulge distribution is equal to the initial (see McLaughlin 1995, Capuzzo-Dolcetta \& Vignola 1996). 
In Table 4 we give the fraction of `lost' clusters for a GCS subjected to dynamical friction and tidal disruption with different nucleus masses and for the two IMFs, with $r_{min}=r_{c0}$. For the  flat IMF, $x$ varies of just 25 per cent at varying the nucleus mass from 10$^7$ $M_\odot$ to 10$^9$ $M_\odot$, indicating that dynamical friction is important. For the less dense clusters of the steep IMF $x$ increases by a factor four over the same $M_n$ range because, in this case, tidal disruption is dominant.
\begin{table}
 \caption{Fraction of `lost' clusters after 15 Gyr.}
 \label{symbols}
 \begin{tabular}{@{}lcc}
  $M_n$   ($M_\odot$)        & s=0   &  s=2  \\
\hline
  10$^7$  & 0.40  & 0.20\\
  10$^8$  & 0.42  & 0.46 \\
  10$^9$  & 0.51  & 0.78 \\
\hline 
\end{tabular}
\end{table}

To understand better the contributions of the two depleting mechanisms and 
their dependence upon  the relevant parameters (nucleus mass and clusters' 
half mass density) we have followed the evolution of the GCS density profiles in two limiting cases: $i)$ pure dynamical friction and $ii)$ pure tidal disruption. The computations were performed on various single-mass GCSs, over a range of individual cluster 
masses $m$, choosing $r_{min}=r_{c0}$. 

In case $i)$, the results at time 15 Gyr are well fitted by a broken power-law,
\BGE
x_{df}=\left \lbrace \matrix{& 0.46\cdot m^{0.82} & 0.01\leq&m&\leq0.6 & \cr
                             & 0.35\cdot m^{0.36} & 0.6 < &m&    < 3    &} \right.
\EDE
where $m$ is in 10$^6 M_\odot$.
For a GCS with a distribution of masses it is quite natural to calculate the quantity $x_{df}$ by averaging it on the IMF. This leads to an 
error of about 3 per cent respect to the exact result obtained by integrating the volume density given by Eq. (10), indicating as reliable the simple average of the expression (16) on different IMFs. For our models, we obtain: for a flat IMF, $x_{df}$= 0.38; for a steep IMF, $x_{df}$=0.04. Clearly the influence of dynamical friction is almost negligible if the GCS's initial mass function is biased towards light clusters.

In case $ii)$, we expect that $x_{tid}$ depends on the nucleus mass and cluster half-mass density through the combination $M_n$/$\sqrt{\bar\rho_h}$. In fact, 
we find that a good fit to $x_{tid}$ is
\BGE
x_{tid}= \left \lbrace \matrix{& 6.79\cdot 10^{-5}y^{0.79}
            &0 \leq y \leq 7.47 \cdot 10^5 & \cr
       &  0.17 \cdot y^{0.21} &  y > 7.47 \cdot 10^5  &} \right.
\EDE
where $y=M_n\cdot \bar \rho_h^{-1/2}$, $\bar \rho_h$ is in units of $M_\odot \cdot$pc$^{-3}$ and $M_n$ in solar masses.
Like before, to obtain the quantity $x_{tid}$ for a GCS with a distribution of masses we average the expression (17) on the IMF. 

When both the effects are at work, a reasonable expression for $x$ would be, clearly:
\BGE
x=x_{df}+\left(1-x_{df}\right)~x_{tid}.
\EDE 
The values obtained in this way are in good agreement with the s=2 case of Table 4, since $x_{df} \ll x_{tid}$, while the s=0 case is poorly represented because the two effects compete.
\subsection{The evolution of the mass function}
Since both dynamical friction and tidal disruption timescales depend differently on the cluster mass, the IMF given by Eq. (13) evolves into a mass function which is different from point to point in the system.

In Fig. 7 we show the present time mass function of the GCS for different nucleus masses (excluding the region within $r_{c0}$ to make it comparable to radially limited observed samples).

Note that, as long as the GCS mass contribution to the nucleus is negligible, the knowledge of the evolution of a particular mass function allows to obtain the evolution of a different IMF. Indeed, under this hypothesis, the individual mass components of the GCS evolve independently of each other. Hence, the following relation among two mass functions $\psi$ 
and $\phi$
\BGE
{\psi_e \over \psi_0}={\phi_e \over \phi_0}
\EDE
holds, where the subscript `0' stands for the IMF and the subscript `e' stands for the evolved mass function.
Thus, it is straightforward to obtain the initial mass function $\phi_0$ from the evolved observed mass function $\phi_e$ via Eq. (19).
Of course, some assumptions on the time evolution of the 
nucleus mass are needed. In a more detailed scenario, the GCS will contribute to M$_n$ in a way dependent on the IMF, which means that Eq. (19) no longer applies.
\begin{figure}
 \vspace{1pt}
\epsfxsize=240 pt
\epsfbox{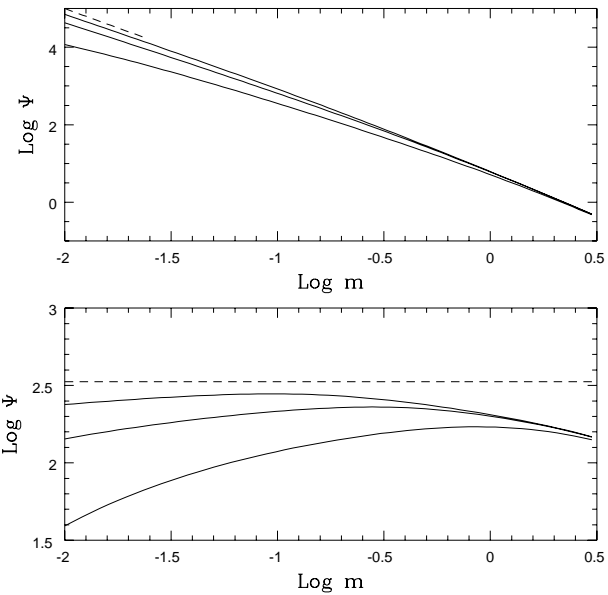} 
 \caption{Evolved (t=15~Gyr) mass functions for different nucleus masses.
Upper panel: steep (s=2) IMF; lower panel: flat (s=0) IMF. The thick solid curve
is the IMF, the other curves correspond to $M_n$=10$^7$, 10$^8$ and 10$^9$ M$_\odot$ (top to
bottom). Cluster masses are in 10$^6 M_\odot$; $\psi$ in units such that $N_{tot}$=$1000$.}
\end{figure} 
\section{Conclusions}
In this paper we have developed a model which allows to follow the evolution of the density distribution of a globular cluster system (GCS) in a triaxial galaxy under the influence of two effects: dynamical friction by field stars and tidal disruption by a massive central object. Both these effects are amplified by the triaxiality of the gravitational potential of the parent galaxy.
The exact knowledge of the present-time density profiles permits to calculate the value of some of the relevant observables as a function of the assumptions made on the initial distribution and mass function of the GCSs. 

Actually, we find that the minor concentration of the GCSs, relatively to the distribution of halo stars, as observed in many galaxies, is an effect which depends 
strongly on the quality of the observations. In particular, a comparison of the core radius of the GCS with that ($r_{c0}$) of its parent galaxy star distribution might be misleading, since the estimate for the core radius of the GCS depends on the minimum radius at which clusters are sampled. For example, we have shown that, in the case of a GCS whose evolution is prevalently ruled by dynamical friction, as it is the case for a GCS made up mostly of heavy clusters, the final shape of the distribution displays a strong concentration of massive clusters in the very inner regions of the galaxy and a consequent lack of clusters in the outer regions. In this case, if the observations do not reach the inner regions (within $r_{c0}$), we effectively measure an erroneusly  large core radius. 
Thus, an observed larger core radius is not a firm signature of the 
presence of a massive object at the centre of the galaxy.
However, we found noticeable differences among the case of a steep IMF and that of a flat IMF. These differences, due to the influence of the half-mass density of the clusters as a function of the cluster mass, are such that:  

$i)$ for a steep IMF, made up mostly of light clusters, the prevailing effect is tidal disruption and there is a strong dependence of the evolved core radius on the mass of the central object (the value $r_c=10 r_{c0}$ is obtained with a nucleus of 2$\cdot 10^8 
M_\odot$); 

$ii)$ for a flat IMF, made up mostly of heavy 
(and dense) clusters, the prevailing cause of depletion is dynamical friction and, consequently, the dependence on the mass of the central object is weak ($r_c= 10 r_{c0}$ for $2\cdot 10^9 M_\odot$).

We have found that a parameter which is more reliable than the core radius to describe the GCS is the number of `lost' clusters outside some minimum radius (we choose $r_{min}=r_{c0}$). This is the difference between the initial and the observed number of clusters, under the assumption that the GCS was initially distributed as the parent halo light. Our calculations show that: for a flat IMF the percentage of `lost' clusters ranges from 40 per cent (with no massive central object) to 50 per cent (for a 10$^9 M_\odot$ nucleus) of the initial total number; for a steep IMF it ranges from 3 per cent (no nucleus) to 80 per cent (10$^9 M_\odot$ nucleus).

We give two formulas which fit the fraction of `lost' clusters in the case of dynamical friction only and in the case of tidal disruption only, as a function of the cluster mass. When both the effects are working, the number of `lost' clusters may not be obtained by simply summing the fractions, confirming that an interaction among the two effects exists.

To conclude, our calculations show that the present differences observed between the GCS and halo stars surface distributions  {\it can be explained by dynamical evolution of the GCS, under the influence of dynamical friction and tidal disruption}, even if the initial concentrations (core radii) were the same. Thus, to have a definite answer to the question "are the observed differences between the star-bulge and the GCS density profiles just reflecting different initial conditions
or are a consequence of evolution?", it would be crucial to compare their kinematical properties. For example, the knowledge of the run with radius of the GCS (projected) velocity dispersion may help to understand if, like in our galaxy, there has been a selective depauperation of clusters on highly radial orbits.

\appendix
\section[]{}
The relation
\BGE
r_a=u^{-1}(E)
\EDE
holds between the orbital energy $E$ and apocentric distance $r_a$, where u(r) is the spherical
symmetric component of the Schwarzschild's potential (see Eq. (5) in Pesce et al. 1992).
Eq. (A1) suffices to give $\tau_{tid}$ as a function of $E$. At any point r, we average the tidal disruption timescale on the distribution function, $f(E)$, of the clusters system, thus obtaining an averaged tidal disruption timescale:
\BGE
\matrix{
&\tau_{tid}(r;M_n,\bar \rho_h)&=& \langle\tau_{tid}(E,M_n,\bar \rho_h)\rangle_{DF}= \cr 
& \cr
& &=&{\int {\tau_{tid}(E,M_n,\bar \rho_h)\cdot f(E)~v^2~dv} \over
\int {f(E)~v^2~dv}}
}
\EDE
where the integration is done on the region of phase-space corresponding to bound orbits.
In the same way, we may obtain an averaged dynamical friction timescale
\BGE
\matrix{
&\tau_{df}(r,m)&=& \langle\tau_{df}(E)\rangle_{DF}= \cr 
& \cr
& &=&{\int {\tau_{df}(E)\cdot f(E)~v^2~dv} \over
\int {f(E)~v^2~dv}}
}
\EDE
\section[]{}
First of all, we seek a solution, say $\tilde\rho$ of the continuity equation in its homogeneous form, that is:
\BGE
{\partial \rho\over \partial t} + {\partial \rho v_r \over \partial r}
+ 2 {\rho v_r \over r} = 0
\EDE
An exact analytical solution is easily found when $v_r(r,t)=u(r)w(t)$, in the following way.

We define the quantity $N(r,t)$ as
\BGE
N(r,t)= 4 \pi \int_0^r {\rho(r',t) r'^2 dr'}
\EDE
Now, integrating Eq. (B1) over the same spherical volume as in Eq. (B2), we obtain an equation for $N(r,t)$:
\BGE
{\partial N \over \partial t} + v_r {\partial N  \over \partial r}= 0.
\EDE
If we define the functions
\BGE
\tau(r)= - \int{ dr \over u(r)}
\EDE
and
\BGE
W(t)= \int{ dt \over w(t)},
\EDE
Eq. (B3) becomes
\BGE
{\partial N \over \partial W} - {\partial N  \over \partial \tau}= 0.
\EDE
whose solution is:
\BGE
N(r,t)=F\left( W(t)+\tau(r)\right)
\EDE
where $F(x)$ is a function constrained by the initial conditions on $\rho(r,t)$.
The solution $\tilde \rho_(r,t)$ is obtained by inverting Eq. (B2):
\BGE
\tilde \rho(r,t)= {1 \over {4 \pi r^2}} \cdot {\partial N \over \partial r} = 
{1 \over {u(r) r^2}} f \left(W(t)+ \tau(r) \right)
\EDE 
Assuming an initial density distribution $\tilde \rho(r,0)$, we find the relation
\BGE
f(x)=\tilde \rho\left(H(x),0\right) u(H(x)) H(x)^2
\EDE
where H(x)$\equiv \tau^{-1}$(x).
Finally, the solution of the homogeneous continuity equation is
\BGE
\matrix{
& \tilde \rho(r,t) &=& {u\left( H(W(t)+\tau(r)) \right) H^2(W(t)+\tau(r))\over
 u(r)r^2} \times \cr
& \cr
& &\times& \tilde \rho \left( H(W(t)+\tau(r)),0\right)      
}
\EDE
For the purposes of this paper, we set w(t)=1 (see Sect. 2), so that W(t)=t and the solution (B10) simplifies to:
\BGE
\matrix{
& \tilde \rho(r,t) &=& {u\left( H(t+\tau(r)) \right) H^2(t+\tau(r))\over
 u(r)r^2} \times \cr
& \cr
& &\times& \tilde \rho \left( H(t+\tau(r)),0\right)      
}
\EDE
When a `sink' term, $S(r,t)$, is present (see Eq. (3)), the solution is in the form
\BGE
\rho(r,t)=\tilde \rho(r,t)\cdot E(r,t)
\EDE
where $E(r,t)$ satisfies
\BGE
{\partial \ln E \over \partial t} + v_r {\partial \ln E  \over \partial r}
+ S(r,t)= 0
\EDE
The solution of (B13) is: 
\BGE
\ln E(r,t) = -\int_0^t{S\left( H(x+\tau(r)) , t-x \right) dx}.
\EDE
Thus, the solution when a sink term is present specifies to:
\BGE
\rho(r,t)=\rho_{om}(r,t) \cdot e^{-\int_0^t{S\left( H(x+\tau(r)) , t-x \right) dx}}
\EDE

\bsp
\label{lastpage}

\end{document}